\documentclass[preprint,journal]{vgtc}       

\ifpdf
  \pdfoutput=1\relax                   
  \usepackage{graphicx}                
  \DeclareGraphicsExtensions{.pdf,.png,.jpg,.jpeg} 
\else
  \usepackage{graphicx}                
  \DeclareGraphicsExtensions{.eps}     
\fi%

\graphicspath{{figures/}{pictures/}{images/}{./}} 

\usepackage{microtype}                 
\PassOptionsToPackage{warn}{textcomp}  
\usepackage{textcomp}                  
\usepackage{mathptmx}                  
\usepackage{times}                     
\usepackage{cite}                      
\usepackage{tabu}                      
\usepackage{booktabs}                  

\usepackage{xargs}                      
\usepackage[pdftex,dvipsnames]{xcolor}  
\usepackage[colorinlistoftodos,prependcaption,textsize=tiny]{todonotes}
\newcommandx{\unsure}[2][1=]{\todo[linecolor=red,backgroundcolor=red!25,bordercolor=red,#1]{#2}}
\newcommandx{\change}[2][1=]{\todo[linecolor=blue,backgroundcolor=blue!25,bordercolor=blue,#1]{#2}}
\newcommandx{\info}[2][1=]{\todo[linecolor=OliveGreen,backgroundcolor=OliveGreen!25,bordercolor=OliveGreen,#1]{#2}}
\newcommandx{\improvement}[2][1=]{\todo[linecolor=Plum,backgroundcolor=Plum!25,bordercolor=Plum,#1]{#2}}
\newcommand{\etal}{\textit{et al.}}
\onlineid{1185}

\vgtccategory{Methodological paper}
\vgtcpapertype{theory}

\title{Animals in Virtual Environments}

\author{Hemal Naik, Renaud Bastien, Nassir Navab and Iain D Couzin}
\authorfooter{
\item
Hemal Naik, Renaud Bastien and Iain Couzin are with Max Planck Institute of Animal Behavior, Konstanz and Centre for the Advanced Study of Collective Behaviour, University of Konstanz and Department of Biology, University of Konstanz, E-mail: {hnaik,rbastien,icouzin}@ab.mpg.de.
\item
Hemal Naik and Nassir Navab are with Technische Universit{\"a}t M{\"u}nchen E-mail: nassir.navab@tum.de. 
}

\shortauthortitle{Biv \MakeLowercase{\textit{et al.}}: Global Illumination for Fun and Profit}

\abstract{
The core idea in an XR (VR/MR/AR) application is to digitally stimulate one or more sensory systems (e.g. visual, auditory, olfactory) of the human user in an interactive way to achieve an immersive experience.
Since the early 2000s biologists have been using Virtual Environments (VE) to investigate the mechanisms of behavior in non-human animals including insects, fish, and mammals.
VEs have become reliable tools for studying vision, cognition, and sensory-motor control in animals.
In turn, the knowledge gained from studying such behaviors can be harnessed by researchers designing biologically inspired robots, smart sensors, and multi-agent artificial intelligence.
VE for animals is becoming a widely used application of XR technology but such applications have not previously been reported in the technical literature related to XR.
Biologists and computer scientists can benefit greatly from deepening interdisciplinary research in this emerging field and together we can develop new methods for conducting fundamental research in behavioral sciences and engineering.
To support our argument we present this review which provides an overview of animal behavior experiments conducted in virtual environments.
} 

\keywords{animal behavior, VR for animals, mechanism of behavior, interactive experiments, closed-loop}

\CCScatlist{ 
 \CCScat{K.6.1}{Management of Computing and Information Systems}%
{Project and People Management}{Life Cycle};
 \CCScat{K.7.m}{The Computing Profession}{Miscellaneous}{Ethics}
}

\teaser{
  \centering
  \includegraphics[width=\linewidth]{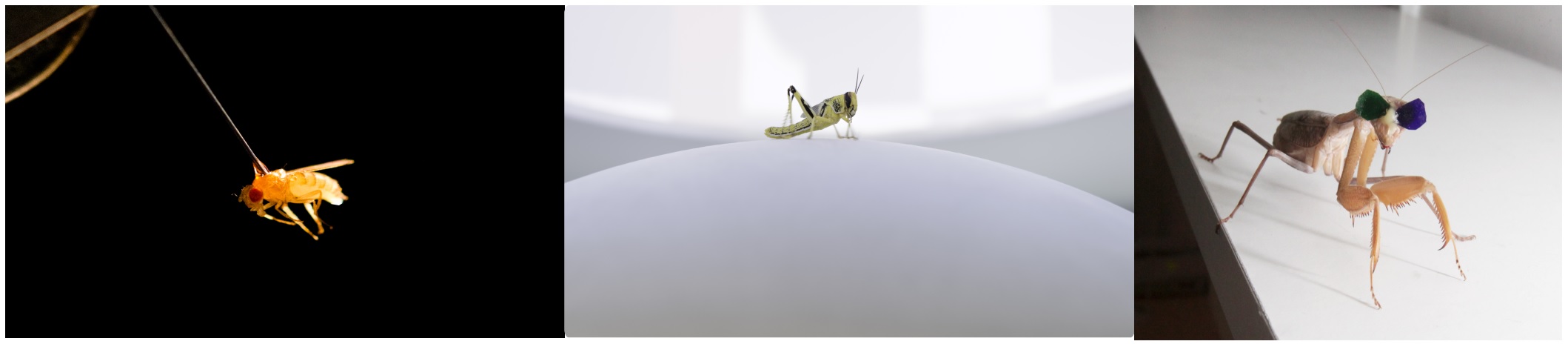}
  \caption{Animals in Virtual Environments: (left) Tethered fruit fly (credit : Simon Gingins), (center) Locust on a spherical treadmill (credit : Centre for the Advanced Study of Collective Behaviour, Konstanz), (right) Praying mantis with glasses \cite{nityananda_insect_2016} (credit : Newcastle University, UK)}
	\label{fig:teaser}
}

\vgtcinsertpkg

\begin{document}


\firstsection{Introduction} 
\maketitle
A wide range of scientific disciplines use animals as a primary subjects of study e.g. medicine, neurobiology, physiology. 
Ethology, the field of animal behavior, is largely concerned with understanding why animals do what they do, and how.
Animals exhibit behavioral strategies that have evolved to enhance its survival in the natural environment (land, air or underwater).
Each animal's behavioral interactions with its own environment, and other organisms, reveals important information about ecology and evolution.
Humans have studied animal behavior for hundreds of years including during domestication. 
In 1963, Niko Tinbergen suggested the first framework for studying behavior in form of four fundamental questions \cite{tinbergen_1963}; What is the survival value of the behavior? How does the behavior develop during the lifetime of the animal? How did the behavior evolve across generations? And how does it work (mechanism)?
His objective was to propose a framework that defines the scope of the scientific study of behavior.
It is widely accepted among behavior researchers that complete understanding of behavior can be obtained from following Tinbergen's framework \cite{tinbergen_1963,bateson_tinbergens_2013}.

Different aspects of animal behavior have been studied over the last 60 years.
In neuroscience, neural activity of behaving animals is recorded to find the link between sensory-motor mechanisms and neural processing \cite{stowers_reverse_2014}.
The genetic basis of behavior can be studied by observing behavior in genetically manipulated animals (such as mutants).
In medicine, small vertebrates (fish or mice) are preferred because they exhibit some fundamental behavioral traits that are consistent with other vertebrates, including humans.
Their behavior can be closely monitored during experimental drug trials to study the progression of the disease and to measure the resulting effect on the animal's behavior \cite{denayer2014_modelSpecies}.
Revealing the behavioral strategies of the animals are useful for solving problems in the fields of engineering and technology.
For example, behavior of animals has been studied for various applications in robotics \cite{arkin1998behavior,taylor_new_2008,jafferis_untethered_2019}.
Biologists and engineers have benefited greatly by working together on novel interdisciplinary projects where robots are used to investigate the principles of decision making in animals \cite{krause_interactive_2011,webb_what_2000,landgraf_robofish:_2016}(details in Sec \ref{Ref:SecDicussion}). 

Animal behavior is studied using various experimental methods.
Behavior is studied in both indoor (lab, cage) and outdoor (wild, open area) environments depending on the research questions.
Outdoor environments are more suitable for the observation of realistic behavioral patterns.
However, experiments in natural environments can be time-consuming and expensive.
Outdoor experiments are also prone to unplanned disturbances from external factors which may alter the behavior during the experiment e.g. weather conditions, human disturbances.
Indoor environments provide more control over the experimental conditions and minimize the influence of external factors.
It is thus easier to develop standardized and repeatable methods for such behavioral experiments.
Indoor environments are suitable for carrying out detailed studies, but the range of behaviors displayed in such environments may be limited. 
Many wild animals do not exhibit natural behaviors in indoor environments.
Often some species, termed as \textit{model species}, are studied in more detail than others because they are selected based on their ability to perform natural behaviors in indoor environments e.g. zebra fish, fruit fly etc.

Artificial sensory stimuli are often used in experiments to invoke behavioral responses from animals. 
In natural conditions, animals constantly receive sensory stimulus (visual, auditory, haptic etc.) from their environment and must react to it appropriately.
Stimuli are often designed artificially to mimic natural conditions.
For example, temperature and light manipulation is sufficient to artificially simulate day and night cycle for insects and birds.
Niko Tinbergen used cardboard models of adult gulls to invoke begging behavior in gull chicks \cite{tinbergen_stimulus_1950}.
Artificial stimulation is a powerful technique for achieving repeatable behavioral observations \cite{Tinbergen_book}.
The experimenter can plan the timing of stimulus delivery and change properties of the stimulus between different trials to observe changes in behavior.
Such experiments provide a deeper understanding of the decision-making of animals.
After Tinbergen's initial findings, more advanced techniques were developed to stimulate sensory systems of different animals for behavioral experiments.
Technological innovations such as cameras, speakers and projectors have made a major impact in behavioral studies.
They are used in novel ways to manipulate the information received by the animal about its surroundings environment (e.g. audio or visual stimulation).

In the late 90s, researchers studying human behavior, psychology, and perception started exploring the Virtual Environments (VE) as a tool for manipulating the human perception of reality by artificial stimulation of human sensing \cite{tarr_virtual_2002}.
The concept of CAVE VR \cite{cruz-neira_cave:_1992} was introduced with the idea of creating an immersive experience for the viewer by means of visual stimulation.
The viewer enters a room where 
head position is tracked and the stimuli are rendered on the walls from the perspective of the viewer. 
Two dimensional figures can appear as three dimensional objects when presented from a specific perspective, an illusion often exploited by graphic artists e.g. M.C.Escher \cite{escher1989escher}.
The CAVE VR was able to create and maintain the illusion in real-time.
Around the same time biologists had shown that the method of displaying virtual stimuli on a screen was useful for studying behavior but it was limited due to lack of interactivity.
Biologists started adopting the CAVE VR design and introduced the concept of interactive virtual environments for animals, almost two decades ago \cite{schuster_virtual-reality_2002,gray_method_2002}.
Their goal was to design a novel experimental approach where the animal behaves as if freely navigating in its natural environment.
Since then many techniques have been developed for studying behavior of freely moving animals (fish, mammals and insects) in the virtual environment e.g. FreemoVR \cite{stowers_virtual_2017}, FlyVR \cite{stowers_reverse_2014}.
Animal VR systems can thus be considered as a cleverly modified version of human VR systems. 

The technology used for designing VE for animals is similar to that used for designing XR applications for human users.
However, the sensory perception of animals is different to that of humans which means that they may sense the environment in a different manner e.g. UV vision in birds, ultrasonic hearing in bats.
The methods developed for stimulating humans may therefore only be suitable for some animals.
Virtual environments for animals are limited by our ability to produce the sensory stimuli that matches the animal's sensory input.
The research in this field has shown that it is possible to circumvent some limitations by exploring new technological solutions i.e. real-time tracking, realistic graphic rendering.
This requires stronger research collaborations between biologists and the technology developers from the XR community.
The goal of this paper is to introduce the XR community to the research done in the field of animal behavior using artificial visual stimuli, especially in virtual environments.
In this review paper, we trace the journey of stimulus-based behavior experiments from simple non-interactive models to fully interactive VR systems.

\section{Review Method and Structure \label{Ref:Sec2RevAndStructure}}
The scientific literature for this review is collected from different research domains associated with the study of animal behavior e.g. ethology, neurology, psychology.
We learned that virtual stimuli have been used extensively in behavioral experiments; therefore, we 
followed a top-down approach to collect the relevant material.
We started from review papers that focused on virtual environments and virtual stimuli for animals \cite{thurley_virtual_2017,stowers_reverse_2014,dolins_technology_2017,dombeck_real_2012,bohil_virtual_2011,leighty_primates_2003,woo_dummies_2011}.
Most of this literature is informative and extensive but it is prepared for readers with backgrounds in biology and behavioral experiments.
Existing review papers discuss the experiments with specific focus on an application (e.g. neurobiology \cite{dombeck_real_2012,bohil_virtual_2011}) or a type of animal (e.g. rodents \cite{thurley_virtual_2017}).
The scope of such a review is restricted by the topic and the methods used for other applications or species are not reported.
Our review provides a more general overview of the methods used to create virtual environments for animals.
We also studied papers with a strong emphasis on the limitations of using virtual stimuli with animals \cite{thurley_virtual_2017,stowers_reverse_2014,chouinard-thuly_technical_2017} and collected literature on non-interactive methods for artificial stimulation.
These methods are commonly used for studying behavior and their success is the strongest argument in favor of developing virtual environments for studying behavior.
Overall our review is designed to serve as a guide for readers of the computer science community (especially XR) who wish to later explore more detailed literature in the field of behavior studies.
This survey will mainly cover experiments using visual stimuli since these represent the largest and most diverse body of work to date (summary of papers in Table \ref{tab:OpenLoop} and Table \ref{tab:ClosedLoop}). 
VEs for other sensory modalities (e.g. olfactory, audio) are relatively few and are mentioned for the sake of completeness.

\paragraph{Structure:}Our first focus will be on experiments explaining a non-interactive (open-loop) approach.
We start with detailing different categories of artificial visual stimuli in the same order as they were introduced in the field of behavioral studies (Sec~\ref{Ref:SecOpenLoop}).
In this way we are able to begin with older experiments which introduced the method of visual stimulation and point out gradual rise in usage of technology in behavioral experiments.
Then we switch our focus to experiments which use an interactive (closed-loop) approach to display different types of artificial visual stimuli, mainly virtual environments. 
We present these experiments in three categories which are based on the novelty of their approach, namely mechanical, hybrid and digital (Sec.~\ref{Ref:Sec5ClosedLoop}).
The advantages and disadvantages of each approach are mentioned with suitable examples.
Following this we discuss the limitations faced by biologists while designing virtual environments for animals 
(Sec~ \ref{Sec:Limitation}).
The technological shortcomings are pointed out to attract the attention of the readers from the computer science community who could contribute important future developments.
Presently behavioral experiments with VEs have reached a stage where support from technology experts is required to design novel techniques for studying complex behavior patterns. 
To support this claim we not only present the current state of the art, but also propose some futuristic ideas for behavioral experiments which can be realized by using techniques from computer vision, XR community (Sec.\ref{Ref:SecDicussion}).

\begin{table*}[h]
	\caption{Overview of artificial visual stimuli used in open loop experiments}
	\label{tab:OpenLoop}
	\scriptsize%
	\centering%
	\begin{tabu} to \textwidth {X[0.5, c]X[1, c ]X[3 , c ]X[2 , c]}
		\toprule
		Type & Stimulus &  Animal-Behavior & Key Attributes \\
		\midrule	
		Static &  Model, Image, Color Filter, Paint conspecifics & Birds - Feeding\cite{tinbergen_stimulus_1950}, Vigilance \cite{evans_use_1991}, Mate choice \cite{bennett_ultraviolet_1996}, social hierarchy \cite{ROHWER_1985_SparrowPattern}. Fish - Mate choice \cite{milinski_female_1990}  & Configurable properties: shape, size etc., reusable method, Non-interactive  \\
		\midrule
		Abstract & Patterns with points, lines, circles & Fruit Fly - Perception and Navigation \cite{schuster_virtual-reality_2002}, Movement of eye\cite{land_eye_2019}, Motion Control \cite{fry_trackfly:_2008}, Trajectory correction \cite{theobald_dynamics_2010}.
		Locust - Motion Parallax \cite{sobel_locusts_1990}, Insect Locomotion \cite{taylor_new_2008}. 
		Moth - Navigation \cite{gray_method_2002}. 
		Mice - Optomotor Response \cite{abdeljalil_optomotor_2005} & Setup can be mechanical design with pattern cylinder or with screens, popular for studying vision induced motion e.g. OMR, OKR \cite{kretschmer_comparison_2017,land_eye_2019,abdeljalil_optomotor_2005}\\
		\midrule
		Video & Video recording displayed through screen or projector & Lizard - Courtship \cite{jenssen_female_1970}, Communication \cite{ord_digital_2002}
		Jumping spider - Recognition \cite{clark_video_1990}, Birds - Alarm calling \cite{evans_use_1991}, Fish - Laterality and Cooperation \cite{bisazza_laterality_1999}.
		Review - \cite{death_can_1998,nelson_use_2013}. & Stimulus can be edited and customized, reusable setup for multiple behavior experiments, can display abstract stimuli, non-interactive. \\
		\midrule
		Virtual & Computer generated content through projector or screen & Fish - Mate preference \cite{baldauf_computer_2009,kunzler_female_2001}, Predator Response \cite{gerlai_zebrafish_2009,Ioannou1212}, Communication \cite{hess_animated_2014}.
		Review - \cite{butkowski_automated_2011,woo_dummies_2011} & Stimulus programmable, reusable setup, semi-interactive or rule based interaction, can display abstract stimulus. \\
		\bottomrule
	\end{tabu}%
\end{table*}

\section{Artificial Stimuli and Open-Loop Experiments \label{Ref:SecOpenLoop}}
In this section we aim to introduce the readers with a brief history of stimulus design and open-loop experiments.
This section is important for readers from non-biological backgrounds to understand how biologists came to use virtual environments for studying behavior of animals. 
First we explain fundamental ideas behind using artificial sensory stimuli and then relate these ideas to the framework of studying behavior \cite{tinbergen_1963}.
We cover the four different categories of visual stimuli that are commonly used for behavioral experiments: static stimuli, abstract stimuli, video stimuli and virtual stimuli.
Each category represents one or more types of visual stimuli which share some common properties in terms of design and/or the method used to present them. 
We describe the intuition behind designing each type and with suitable examples we show how these stimulus were implemented.
In the end the knowledge gained from these experiments is summarised.

In these studies, researchers wanted to design an efficient method to provoke meaningful behavioral response for scientific analysis.
Every novel idea of visual stimulation was first tested with an open-loop (non-interactive) approach.
In such an approach the stimulus does not change or react to the animal but the method is sufficient to check the feasibility of using a stimuli and to learn the right technique for displaying it.
Experiments with the open-loop approach are still being used in behavioral studies and they have provided much of the fundamental understanding required to build the modern closed-loop behavior experiments.

\begin{figure}[tb]
 \centering 
 \includegraphics[width=\columnwidth, height=6cm, keepaspectratio]{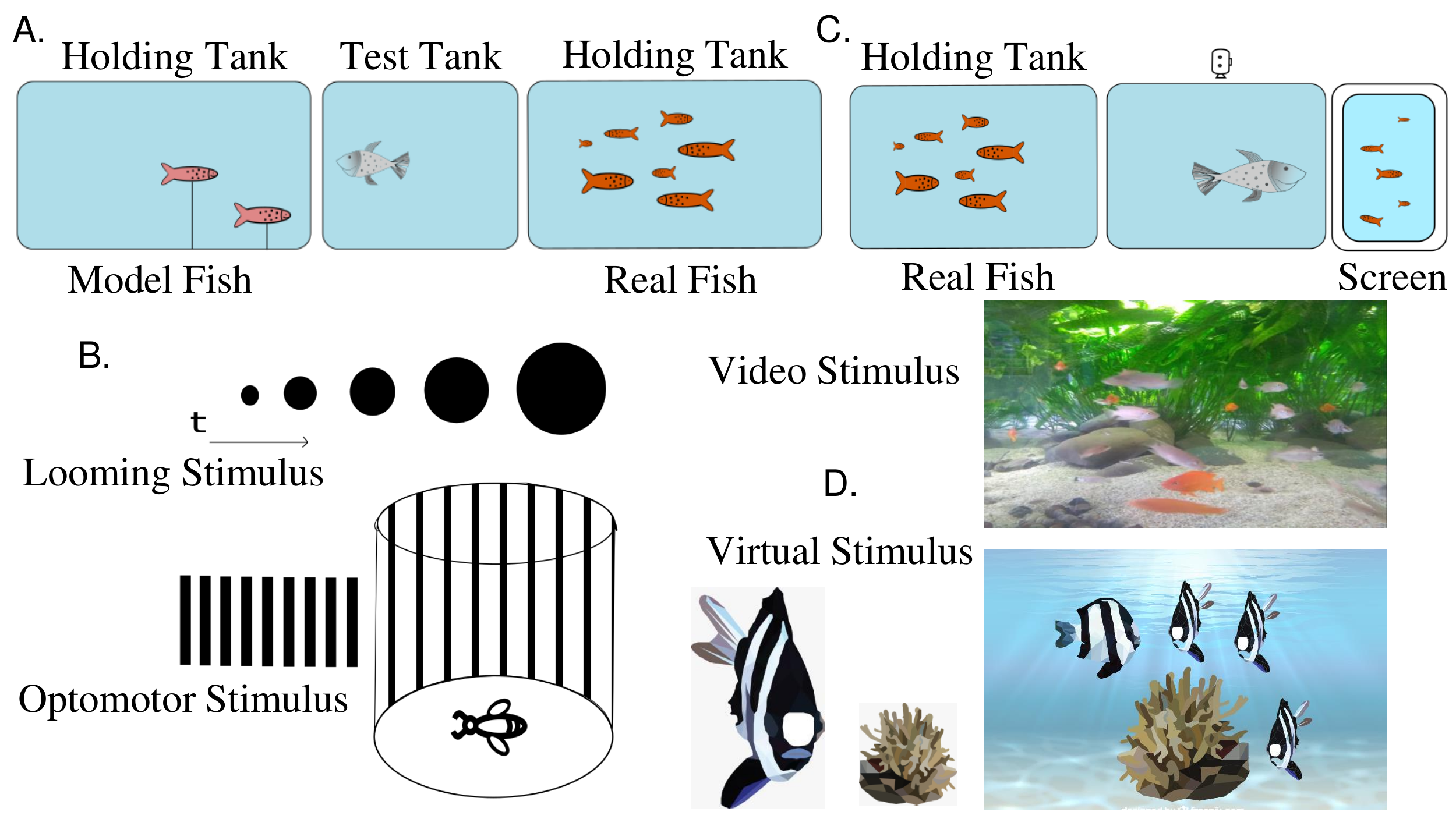}
 \caption{Common type of visual stimuli used in open-loop behavior experiments. A. Static stimuli: An experimental setup designed to study preference of the fish in the test tank when given a choice between model fish and real fish, B. Abstract stimuli: Common type of patterns used to study vision induced motion and pictorial depiction of mechanically controlled Pattern Cylinder (PC) with striped patterns, C. Video stimuli: Experimental setup for preference study with video playback experiment, similar to figure A., the screen is used to display videos fish as shown in example image, D. Virtual stimuli: Example of an image displayed as virtual stimulus. It consists of some fish around a coral, all of which are separately designed graphical models (left) combined with an underwater background. This stimulus is displayed using a screen similar to the setup shown in figure C.}
 \label{fig:StimulusTypes}
\end{figure}

\subsection{Static stimuli}
Static stimuli were some of the first employed in behavioral experiments. 
Here animals are presented with a static object such as a model or an image, usually of an animal (figure \ref{fig:StimulusTypes}A).
We refer to this type as static because the properties of stimuli do not alter or change during the experiment. 
It was hypothesized that animals may perceive a static model as a real animal and react to it. 
It was shown that in some cases such stimuli were sufficient to invoke a response such as fear or attraction. 
Tinbergen and Perdeck \cite{tinbergen_stimulus_1950} used a model of a bird to invoke begging behavior in chicks of gulls. 
The chicks responded naturally to the models i.e. as if begging for food from a parent.
Following such studies, other researchers tested artificial objects extensively using different variants, which differed both in terms of visual properties of the model (e.g. different colors, shapes), and timing of stimulus delivery (e.g. time of the day or frequency) \cite{death_can_1998}.
Evans and Marler \cite{evans_use_1991} studied alarm behavior in chickens using a model of a predator.
They created a setup where a model of a predatory bird would move above the cage on a rope. 
This simulated a typical behavior of a predatory bird gliding in the sky looking for food.
It was observed that chickens made alarm calls when the model was moving over the cage.
Images, photographs and slides were also used as static stimuli \cite{death_can_1998}.
Other examples of static stimuli are environmental modifications e.g. light filters, which allow or reject of specific wavelengths \cite{milinski_female_1990,bennett_ultraviolet_1996}, and visual modification e.g. painting conspecifics \cite{ROHWER_1985_SparrowPattern}.
Robotic animals are increasingly used as visual stimuli \cite{landgraf_robofish:_2016,klein_robots_2012,webb_what_2000,krause_interactive_2011}, but to maintain focus on virtual visual stimuli we do not cover them in this paper.

Static stimuli have proved to be reliable and repeatable means for conducting behavioral experiments.
The main advantage is that the same stimulus could be used for different animals and its visual properties could be modified between trials.  
Moreover, the timing of stimulus delivery and frequency of displaying the stimulus could also be controlled \cite{death_can_1998}. 
A major limitation to this approach is that there is no scope for feedback between the stimuli and the organisms. 
Consequently a problem that arises is that individuals can habituate to, and stop responding to, stimuli over time \cite{death_can_1998}.

\subsection{Abstract stimuli \label{subSec:Abstract}}
The primary intuition for the design of abstract stimuli was to design a stimulus that is minimalistic yet sufficient to drive behavioral decision-making process in animals (such as movement).
These are the most widely used stimuli for behavior experiments, especially for studying mechanisms related to visually-induced locomotion.
For example, a common abstract stimulus consists of simple patterns designed from primitive geometric shapes such as points and lines (see figure \ref{fig:StimulusTypes}B) e.g. stripes or circles.
The idea is to measure the movement of the animal in response to the patterns displayed to it.
The mapping between the features observed by the animal and the animal's movements reveal the underlying process of behavioral decision making in the context of such stimuli.
This experimental concept is designed to investigate fundamental questions regarding the behavioral and neural basis of visually-induced locomotion.
Small invertebrates, such as fruit flies or honey bees, and relatively simple vertebrates such as fish, are the preferred model species as they possess less complex nervous systems and relatively fewer behavior patterns.

One example of visually-induced locomotion behavior is the optomotor response (OMR), which is the property of moving the body and/or head in concert with the features in the environment for image stabilization \cite{kretschmer_comparison_2017,land_eye_2019}.
Similarly, the property of moving the eyes in concert moving features is called optokinetic response (OKR).
One of the first experimental setups for studying OMR and OKR was a mechanically controlled stimulus delivery system.
It consists of a stage for placing the animal and a cylindrical drum surrounding the stage.
The cylinder is rotated along its axis using a motor and its inner walls are painted with abstract patterns (stimulus) cf. figure \ref{fig:StimulusTypes}B.
The movements of the animals are restricted to a small area and sometimes tethering is used to keep the animals fixed in one spot which simplified measurement of head or eye movements.
The movement is recorded using video cameras or using a simple array of photodiodes \cite{gray_method_2002} or torque motor \cite{geiger_optomotor_1974}.
It is shown that the animals typically display a tendency to move in the direction of the rotation.
The width of the striped pattern, the rotation speed of the cylinder, and the direction of the rotation are also influential in decision making \cite{abdeljalil_optomotor_2005}.

Another example of visually induced motion is avoidance of or flight response to, a rapidly-expanding (or ‘looming’) shape on the retina, typically a black expanding circle with a white background as shown in figure \ref{fig:StimulusTypes}B.
In early designs, looming stimuli were simulated by mechanically moving a dark circular cardboard cutout closer to the animal.
Electronically controlled display methods (e.g. LED grid, LCD) have replaced the mechanical methods for displaying the stimulus and computer vision techniques are now deployed for movement measurements \cite{schuster_virtual-reality_2002, strauss_processing_1997}. 
It is also possible to conduct abstract stimuli-based experiments in a fully automated closed-loop manner \cite{fry_tracking_2000,stowers_reverse_2014,fry_trackfly:_2008} (covered later in virtual stimuli).

Experiments with abstract stimuli are relatively easy to design and their results tend to be reproducible and the method has opened doors for reverse-engineering the process of visually induced motion.
Abstract stimuli have been successfully used for more than 70 years for research studies by biologists and engineers alike.  
Some examples include studying the vision properties such as depth perception \cite{schuster_virtual-reality_2002}, motion parallax \cite{sobel_locusts_1990}, movement of the eye \cite{land_eye_2019}, physiology of the eye \cite{borst_drosophilas_2009}, locomotion mechanisms such as flight behavior \cite{taylor_new_2008} and motion control \cite{fry_trackfly:_2008}, and behavioral patterns like navigation \cite{gray_method_2002,schuster_virtual-reality_2002}, and trajectory correction \cite{theobald_dynamics_2010}. 
The experiments have been conducted exhaustively with small insects (e.g. fruit flies, bees) and vertebrates (e.g. zebrafish, mice).
Mechanical designs have suffered from mechanical limitations and the ability to measure movement.
They also require manual intervention for changing patterns or other parameters, which not suitable for conducting high-throughput experiments.

\subsection{Video stimuli}
This category includes experiments in which recorded video footage is used as visual stimuli.
These experiments are known as video playback experiments in the behavior literature.
For experiments investigating social behavior video cameras are typically used to record the activity of an animal and this is then used as a stimulus during the experiment \cite{death_can_1998,nelson_use_2013}. 
The focal animal is usually placed in an enclosure or an arena and is shown a video sequence through a screen or projector placed at a reasonable distance (figure \ref{fig:StimulusTypes} C).
The responses of this animal is recorded using a video camera, which are later used to map behavioral decisions of the animals to the visuals presented in the stimuli.
The stimuli may involve another animal of the same species (conspecific) or a different species (heterospecific) behaving in a certain way.
It was hypothesized that animals do not comprehend the concept of video screens and thus will react in a natural and instinctive manner to the presented stimuli.
The intuition behind this method is to simulate the natural environment in the photo realistic way \cite{death_can_1998}.

Jenssen \cite{jenssen_female_1970} designed a mate-choice experiment with female lizards using video playback method.
He displayed video sequences of male lizards performing courtship display and reported that female lizards did react appropriately to such stimuli.
The video playback technique is commonly used method for studying a wide range of behavior responses such as: aggression, attraction, fear etc, in species including arachnids \cite{clark_video_1990}, birds\cite{evans_use_1991}, reptiles \cite{ord_digital_2002}, and fishes \cite{bisazza_laterality_1999}.
It is a reliable technique for quantitative analysis of behavior.
The movements are measured in 2D or 3D space using multiple cameras \cite{pitcher_simple_1984}.
Playback methods have provided many insights into the questions related to the survival value of a specific behavior. 

Video stimuli have some clear advantages because a customized sequence of behaviors can be shown.
The same setup can be used to display wide range of behaviors; for example the setup in figure \ref{fig:StimulusTypes}C can be used for studying either mate preference or aggression.  
Camera and display technologies required for the experiment are typically commercially available which makes this method accessible to researchers.
Various software tools are designed to quantify behavioral response from the video sequences \cite{dell_automated_2014}.
Detailed behavioral studies are possible because the entire stimulus sequence is known and it could be mapped to the response of the animal in an offline manner.

Video playbacks methods also have many disadvantages.  
They often assume that the animal is reacting to the stimulus and perceive the stimulus as being real.
This assumption may not always hold. 
Stimuli are customized yet they are mostly pre-programmed sequences \cite{death_can_1998}.
The animal in the videos do not interact fully with the real animal which may lead to habituation or unnatural reaction.
Other limitations are introduced with the use of display and video technology.
The sampling rate of the camera used to create the stimulus must match the physiological visual properties of the animal otherwise the motion in the video may appear blurry, discolored or distorted to the animal.
Technical specification of the display screen or projector must also be considered to avoid similar problems e.g. resolution of the display, refresh rate of screens, etc.
There are some common limitations shared by all screen-based methods of visual stimulation which are covered in detail later in section \ref{Sec:Limitation}.
Video playback methods are limited to those behaviors which are possible to record.
Further details on feasibility of using video stimuli can be found in the review paper of D'eath \cite{death_can_1998}.
In summary, video playback methods made a strong case in favor of using technology provided the researcher considers carefully its limitations and makes reasonable assumptions.

\subsection{Virtual stimuli}
Virtual stimuli are the most advanced category of artificial visual stimuli.
The setup is more or less similar to video playback methods but the content of the stimulus is created virtually i.e. using computer graphics and animation technology.
The initial motivation for using virtual methods was to increase immersion and remove limitations of earlier methods.
In a computer-generated stimuli, the user can configure fine details, which is not typically possible with raw video stimuli (see figure \ref{fig:StimulusTypes}D).
All components of the stimuli are programmable and can be modified independently i.e. the shape, size, color, background, and movements of the animals can be individually changed for each experiment.
Graphic design and rendering are performed using techniques developed by the video game and animation industry.

Virtual stimuli are considered a major improvement over other stimulus types. 
The stimuli are faster to design and modify which means multiple experiments can be conducted with different variants.
Video playback and abstract stimuli-based experiments were also performed with virtual stimuli for cross-validation \cite{woo_dummies_2011}. 
Virtual stimuli were proposed as an alternative to other screen-based methods and this method has been widely adopted for studying different behavior patterns (fish) e.g. mate preference \cite{kunzler_female_2001,baldauf_computer_2009}, predator response \cite{gerlai_zebrafish_2009}, and visual communication \cite{hess_animated_2014}.
New methods have been employed to create realistic animals and scenes.
Ioannou \etal\cite{Ioannou1212} projected small dots onto a surface of a fish tank to simulate the movement of very small prey.
The fishes attacked the projections as if they were real prey, which helped the authors to understand the hunting strategy of the fish, as well as to conduct artificial evolution of the prey.  
anyFish \cite{veen_anyfish:_2013} and FishSim \cite{muller_virtual_2017} are software packages to simulate 2D projection of a 3D animated fish.
Joysticks are employed to define the motion of the virtual animal in semi-interactive manner, or to introduce perturbations in the stimuli \cite{muller_virtual_2017,leighty_primates_2003}.
Abstract stimuli-based experiments benefited significantly from digitization.
Software packages have been designed to automate the workflow i.e. stimulus delivery, behavior measurement (locomotion) and data analysis.
Fry \etal\cite{fry_trackfly:_2008} designed fully automated setup for open-loop experiments with abstract patterns. 
They used an optical tracking method for computing 3D trajectory of a freely flying fruit fly in real-time.
Most importantly, the combination of virtual stimuli and real-time tracking methods provides an opportunity to design closed-loop experiments.
We cover closed-loop methods with virtual stimuli separately in the next section.  

Virtual stimuli-based methods have many advantages when compared with previously described categories.
The modular approach of software is an advantage as it allows different modules related to display, rendering or measurement (tracking) to be changed as and when the new versions are developed.
The software is easier to distribute and share with other scientists in the community. 
The stimulus itself can be programmatically controlled which was not the case with any other method.
However, display technologies are usually made for the human visual system and may not be sufficient to reproduce a realistic view of the animals i.e. they may lack sufficient spatial resolution, spectral resolution.
Technological problems are inherent in virtual stimuli-based methods and they are discussed in detail with limitations of the closed-loop experiments.
The open-loop experiments with virtual stimuli are restrictive as they lack interaction.
For that reason repeated trials with the same animals were not advised as they got used to the stimuli \cite{Ioannou1212}. 
Butkowski \etal \cite{butkowski_automated_2011} and Woo and Rieucau \cite{woo_dummies_2011} reviewed use of animation for open-loop behavior experiments. 

\begin{table*}[tb]
	\caption{Overview of artificial visual stimuli used in closed loop experiments}
	\label{tab:ClosedLoop}
	\scriptsize%
	\centering%
	\begin{tabu} to \textwidth {X[0.5,c]X[1,c]X[1,c]X[3,c]X[2,c]}
		\toprule
		Type & Design & Feedback method &  Animal-Behavior & Key Attributes \\
		\midrule	
		Mechanical &  Arena, Pattern cylinder & Torque meter, Treadmill & Fruit Fly - Pattern recognition \cite{dill_visual_1993}, Motion perception \cite{bulthoff1982drosophila} & Motion based rotation of pattern cylinder, features not configurable. \\
		\midrule
		Hybrid & Arena, LED Screen, Projection & Optical Sensor, Photodiode, Optical Tracking, Treadmill & Fruit fly - Depth Perception\cite{schuster_virtual-reality_2002}, Moth - Neurophysiology \cite{gray_method_2002},
		Rodent - Navigation \cite{madhavplace2015}, Neural activities \cite{harvey_intracellular_2009,dombeck_functional_2010,dombeck_imaging_2007,domnisoru_membrane_2013}, Review - Neuroscience \cite{bohil_virtual_2011,dombeck_real_2012}, Primate Cognition \cite{dolins_technology_2017}, VR for animals \cite{stowers_reverse_2014}, Rodent - \cite{thurley_virtual_2017}  & Animals are restricted or tethered, fully interactive, built with open source software frameworks, setup configurable for multiple species.  \\
		\midrule
		Digital & Arena, Projection & Optical Tracking, Active treadmill & Fruit fly - Real-time 3D tracking \cite{fry_trackfly:_2008}, vision induced motion \cite{stowers_reverse_2014},  Flight pattern \cite{stowers_virtual_2017} 
		Spider - Navigation \cite{ord_digital_2002,stowers_reverse_2014}, Ant - Foraging \cite{dahmen_naturalistic_2017}, Fish - Social behavior \cite{stowers_virtual_2017}, Rodent- \cite{del_grosso_virtual_2017,stowers_virtual_2017}, Review - Neuroscience \cite{bohil_virtual_2011,dombeck_real_2012}, VR for animals \cite{stowers_reverse_2014}, Rodent - \cite{thurley_virtual_2017} & Free moving animals, real time perspective correction, underwater projection, arbitrary surfaced arena, support for multiple species, configurable software.\\
		\bottomrule
	\end{tabu}%
\end{table*}

\section{Closed loop Experiments and Virtual Environments \label{Ref:Sec5ClosedLoop}}
In this section, we focus on behavioral experiments using artificial visual stimuli in a closed-loop.
When navigating in a three-dimensional environment, the features visible to the eye change appropriately with perspective and movement of the viewer.
The fundamental idea of a closed-loop experiment is to constantly update the visual stimuli according to the movement of the animal.
This design has two major components, \textit{tracking} and \textit{stimulus delivery}.
It is necessary to synchronize these two components in real-time for a realistic appearance.
Real-time tracking and perspective correction for a freely moving animal is a difficult problem.
Over the past two decades, different techniques have been developed to circumvent this problem mostly by restricting the movement of the animal. 
Behavioral experiments with virtual stimuli are often referred to as Virtual Environments (VE) or Virtual Reality (VR) interchangeably in the behavior literature.
For the sake of clarity, we will use the term VE generally for closed-loop experiments with virtual stimuli.
We reserve the term VR (as a subset of VE) specifically for closed-loop experiments where the position of the animal is tracked or the animal is maintained stationary to render the stimuli in a perspective correct manner.
Based on the design of the experimental setup we have divided closed-loop experiments into three categories: mechanical design, hybrid design and digital design. 

\begin{figure}[tb]
 \centering 
 \includegraphics[width=\columnwidth]{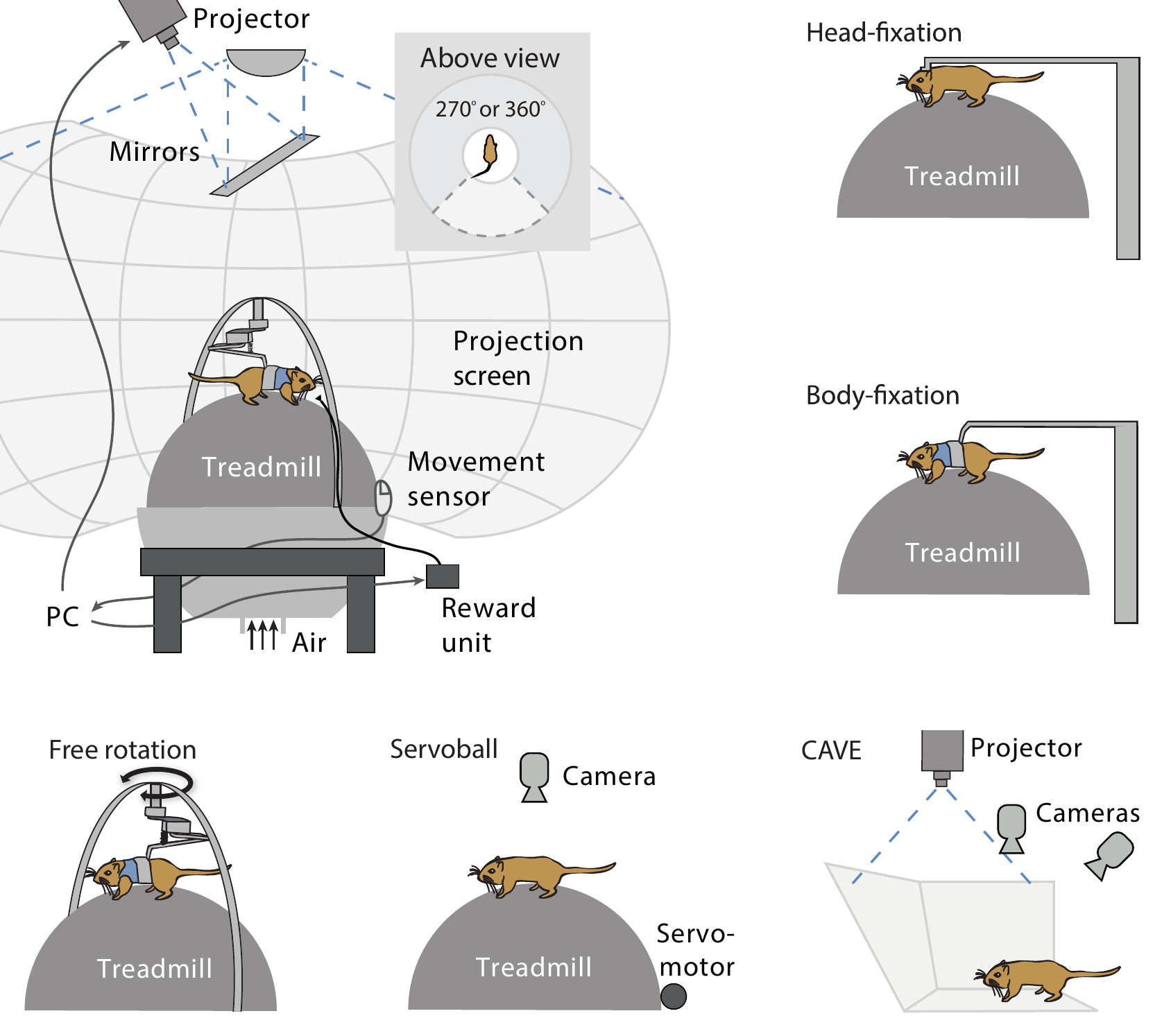}
 \caption{Examples of different type of VR systems (credit : cf. Thurley and Ayaz \cite{thurley_virtual_2017}). Figure A-F show different techniques used for fixation of animals, recording of movement and display of stimulus. Details covered in text.}
 \label{fig:ClosedLoopVRParadigm}
\end{figure}

The idea of VE for animals is largely inspired from the CAVE VR \cite{cruz-neira_cave:_1992} setup designed for humans.
The setups designed for animals are similar and have remained so from last two decades.
An enclosure is designed and the animal is placed on the stage or a platform.
The platform is surrounded by a screen often toroidal or cylindrical, preferably matching the field of view of the animal, for displaying the stimulus e.g. figure \ref{fig:ClosedLoopVRParadigm}A\cite{thurley_virtual_2017}.
VEs are used for numerous behavioral experiments but the experiments mentioned in this section are selected based on the novelty of the approach.
The focus is to highlight critical improvements in behavior experiments using virtual environments.
We show that many of these improvements are largely dependent on the methods developed for computer vision and XR applications.

\subsection{Mechanical design: Restricted animal in fixed (non-virtual) environment} 
One of the earliest designs of a closed-loop experiment was a mechanically designed flight simulator for insects.
The setup is similar to the rotating pattern cylinder design explained earlier in section \ref{subSec:Abstract}.
In the flight simulator the rotation of the pattern cylinder is coupled to the motion of the insect via a torque motor.
This way the motion of the insect triggers the rotation of the cylinder in its visual field, which emulates a real-life flight conditions. 
Dill \etal \cite{dill_visual_1993} used a this setup to study visual pattern recognition in fruit flies and showed that flies could remember patterns based on how they appear on the retina from a specific perspective.
The fly was tethered and its head rotation was immobilized to restrict the movement of head independent of thorax.
The turning response was recorded by measuring the movement of thorax. 
Often head fixation is used to force the insect to turn its body instead of the head.
Tethering is used to control the sensory experience and it simplifies the tracking problem by restricting the movement of the animal and allows experimental recordings to be made during the experiment.

A treadmill with a styrofoam ball is another variation of a mechanically designed closed-loop experiment.
In this case, a tethered animal is placed on the ball and walking motion of the animal rotates the ball which in turn rotates the pattern cylinder.
The rotation of ball is converted to electronic signals which serve as input to servo motor for rotating the cylinder.
The ball is painted with a pattern of infrared reflective dots.
An infrared LED is placed near the ball and its reflection is picked up by a photodiode which further decomposed the rotation and translation using sequence detector logic.
The mapping between the animal's movements and the visual pattern is stored in the computer for further analysis.
B\"ulthhoff \cite{bulthoff1982drosophila} used this setup to study the genetic link between vision and motion perception in fruit flies.
He used genetically modified flies (mutants) and wild fruit flies in the flight simulator to perform a navigation task. 
He showed that mutants showed defect in visual orientation and therefore concluded that optomotor response may be encoded in genetic experession of the animal.

Mechanically designed closed-loop experiments were mainly used to study visually induced motion with abstract stimuli.
These experiments did demonstrate that animals show a preference for some patterns and actively move towards their preferred pattern.
However, the patterns remained unchanged or fixed during the experiments which was considered a major limitation of this approach. 
This limitation is alleviated in the modern closed-loop designs which use projectors instead of pattern cylinders.
The concepts of tethering and treadmills in the modern VE setups are adopted from the flight simulator experiments.

\subsection{Hybrid design: Restricted animals in VE }
Closed-loop experiments with the hybrid design were motivated by the success of virtual stimuli in open-loop experiments.
In hybrid designs, virtual stimuli are displayed using screens or projectors (instead of rotating pattern cylinder), and treadmills and/or tethering techniques are used to restrict the animal’s movements and to simplify the problem of tracking (cf. figure \ref{fig:ClosedLoopVRParadigm}).
In the late 90s, the behavior researchers started the development of closed-loop experiments with virtual stimuli. 
It was easier to configure virtual stimuli to show desired patterns and the appropriate display technologies started becoming commercially available at the time.
Treadmills based techniques were readily available and useful for precise perspective correction while rendering the stimuli on the screen.
These experiments are often referred to as the first experiments which put animals in Virtual Environments. 
VE allowed researchers to try different types of stimuli which extended the scope of research to other behaviors in the three-dimensional world i.e. navigation, foraging, etc.

Schuster~\etal\cite{schuster_virtual-reality_2002} designed one of the first experiments with fruit flies in VE. 
The fly was placed on a stage surrounded by a 360\textdegree  panoramic LED screen which displayed the stimulus pattern.
The 2D movement of a walking fly was measured in the X-Y plane using a simple computer vision technique of blob detection. 
The fly was tethered and its wings were clipped to restrict its movement to a plane.
The authors claimed that they were able to study depth perception in fruit flies with the system which was not possible in previously designed open-loop methods. 

Another notable approach is from Gray \etal\cite{gray_method_2002}, where they designed a VE to measure neurophysiological activity in moths while foraging.
Moths navigate in a complex 3D environment to find the source of odor and the authors simulated similar conditions in VE by designing a multisensory stimulation (visual, olfactory and mechanosensory) mechanism.
A wind tunnel was placed in front of the moth for olfactory and mechanosensory stimulus.
A 3D scene of a textured floor and vertical pillars was generated using computer game engine (Descent III).
The moth was tethered and multichannel neural recording was obtained by probing the ventral nerve cord of the moth.
It was assumed that flight is at a constant altitude (fixed Z) and navigation was allowed for in the simulated horizontal plane.
The movement of the moth's abdomen was measured using optical sensors; an Infrared (IR) light source and photodiode array.
It was shown that the moth navigated in the virtual space by turning towards the odor emanating from the wind tunnel.
The authors demonstrated effectiveness of virtual flight simulator by showing that the turning sequence in VE was consistent with findings of optomotor response observed with freely flying moths.
Generally, fixation of the animal is considered a limitation of this approach. The researchers studying the neural link between stimuli and behavior preferred fixation of animals to be able to measure the neural activity in a reliable manner \cite{thurley_virtual_2017, stowers_reverse_2014}.  

Numerous variations of treadmill-based designs have been used to study the movement of rodents in virtual environments (cf. figure \ref{fig:ClosedLoopVRParadigm}). 
Each technique has imposed different degrees of constraint on the movement of the animal e.g. body fixation, head fixation, etc. 
H{\"o}lscher \etal \cite{holscher_rats_2005} designed the first VE setup for rodents, similar to figure \ref{fig:ClosedLoopVRParadigm}A.
The movement of the treadmill was restricted to the horizontal axis, the body of the animal was fixed but the head position was not.
Rotation of the treadmill was computed using optical sensors, similar to tracking the ball in the computer mouse. 
Visual stimuli were projected using a DMD projector on a screen, via two reflective mirrors to cover wide field of view (360\textdegree ~azimuth and -20\textdegree to +60\textdegree~elevation).
The VE contained cylinders hanging downwards from the ceiling and no features on the floor.
This design was meant to avoid giving any tactile feedback that the rat may expect from visual cues.
OpenGLPerformer was used to generate graphics with support of NVIDIA graphic card for real-time rendering. 
The stimulus was presented at a fixed distance and stereoscopic depth cues were not considered. 
They showed that rodents could be trained to navigate in a virtual 3D environment using 2D stimuli, which had been shown in primates and humans.
The rats were trained for a real maze navigation task to compare their ability to learn in the real world and in VE.
It was shown that they learned to operate the treadmill to navigate "closer" to the objective in VE to earn a food reward.
They got better with the number of attempts and consistently minimized the distance to, and thus the time taken to reach, the reward.
This method was a significant improvement to classical lab experiments where actual mazes must be constructed in order to study navigation.
Restriction of movement and lack of other stimuli such as vestibular, tactile or olfactory cues was considered a major limitation of this approach.
However, the ability of rats to learn and adapt to a new environment while suppressing lack of information from other sensory inputs, was nevertheless considered positively for the use of virtual environments.
Head fixation and body-fixation techniques (cf. figure \ref{fig:ClosedLoopVRParadigm}B,C) were used for head stabilization during measurement of neural activity.
The techniques included recording membrane potentials \cite{harvey_intracellular_2009}, two photon microscopy \cite{dombeck_imaging_2007}, two photon calcium imaging \cite{dombeck_functional_2010}, patch-clamp recording \cite{domnisoru_membrane_2013}.  

In most cases, VE designs are modified to match the visual properties or motion properties of the animals.
Modifications are necessary to answer species-specific questions and improve the realistic appeal of the stimulus.
Free motion treadmills (cf. figure \ref{fig:ClosedLoopVRParadigm}D), for example, were designed to introduce vestibular information about rotation, which was missing in the earlier designs \cite{madhavplace2015}.
Takalo \etal \cite{takalo_fast_2012} used a large field of view and increased the temporal resolution to render stimuli for fast-moving American cockroaches (tethered).
Dahman \etal \cite{dahmen_naturalistic_2017} used hollowed styrofoam design for accurate registration walking speed of desert ants (tethered).
They showed that ants changed their pace significantly between different approach and search phases, they slowed down significantly while approaching nest position.
Stowers \etal \cite{stowers_reverse_2014} designed visual stimuli with configurable chromatic properties to increase the naturalistic appearance of the scene for jumping spiders. 
The authors claim that such systems are well suited to study visual features important for decision-making behavior such as target selection or predator avoidance.

\subsection{Digital design: Free moving animals in VR}
\begin{figure}[t]
 \centering 
 \includegraphics[width=\columnwidth]{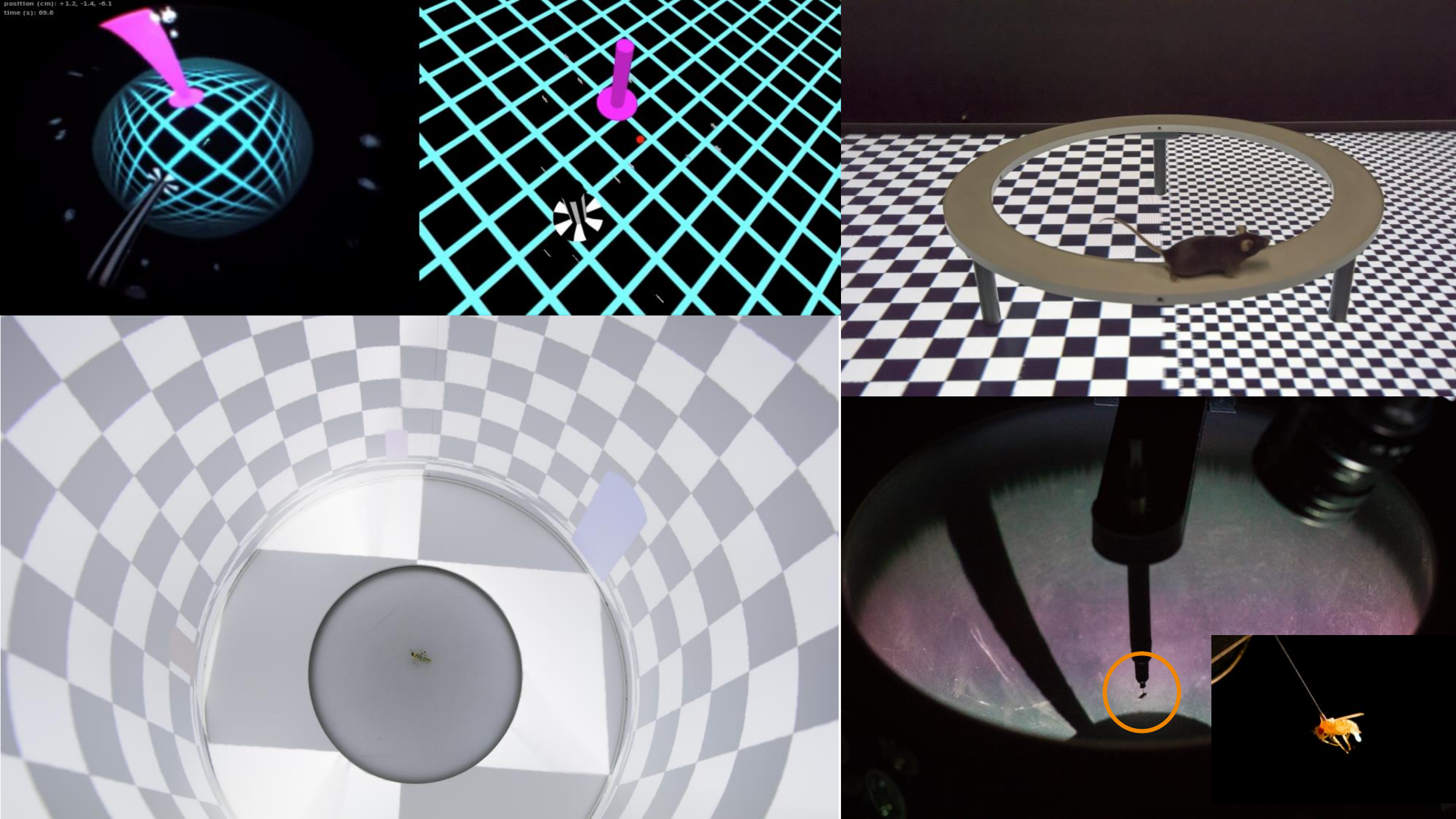}
 \caption{(clock-wise) Example of stimuli from the fishVR system \cite{stowers_virtual_2017} rendered from perspective of fish (left) and human (right), MouseVR system with free moving rat on circular platform \cite{stowers_virtual_2017} (credit: https://strawlab.org/freemovr), FlyVR setup with tethered fly (credit: Simon Gingins), top view of VR arena made for terrestrial insects (credit : Centre for the Advanced Study of Collective Behaviour, Konstanz ) }
 \label{fig:FreemovingVR}
\end{figure}

In this subsection, we cover experiments with true Virtual Reality designs which allow free movement of the animal.
This means that the stimulus is rendered in such a way that it creates an illusion of a three-dimensional space from the animal's perspective, even though the projections are on a 2D surface at a fixed distance.
This compensation of view is known as perspective correction in human XR literature. 
Perspective correction is achieved using sophisticated computer vision techniques of employing real-time tracking (with multiple cameras) of the animal's head in 3D. 
The tracking data is provided to the rendering engine with a minimum delay to provide real-time projection, correctly rendered from the perspective of the animal as it moves through the virtual space. 
The stimulus can be displayed on flat or arbitrary shaped surfaces (see figure \ref{fig:FreemovingVR}) using multiple projectors or screens operating at high framerate.
The existing systems use advanced concepts from computer vision and XR research such as multi camera-projector synchronization and calibration, real-time 3D tracking and rendering \cite{fry_trackfly:_2008,stowers_reverse_2014}.  

One method to design VR systems with freely moving animals is to use an active treadmill.
Active treadmills are used to compensate motion of the animal and keep them in stationary position (cf. figure \ref{fig:ClosedLoopVRParadigm}E,\ref{fig:FreemovingVR}).
The animal is tracked in 3D using a video camera and counter-motion of a treadmill is triggered through a servo motor to keep the animal in the same physical location.
This type of design facilitates behavior studies without creating very large arenas e.g. navigation behavior in jumping spider \cite{peckmezian_virtual_2015} \cite{stowers_reverse_2014} or foraging behavior in desert ants \cite{dahmen_naturalistic_2017}.
Treadmill-based solutions are not suitable for all animals and therefore the development of novel 3D tracking methods was crucial for the development of VR solutions for animals.
Fry \etal\cite{fry_trackfly:_2008,fry_tracking_2000} created TrackFly framework to conduct high throughput closed-loop experiments with unrestrained flying fruit flies. They tracked free moving flies at 50~Hz using a multi-camera setup. 
Building upon this tracking approach Stowers \etal \cite{stowers_reverse_2014} built a FlyVR system.
Markerless tracking was done with infrared filters to facilitate fast image processing and block the visible light from stimulus screens.
Stowers el at.\cite{stowers_reverse_2014} showed that it was possible to combine real-time 3D tracking and stimulus delivery to induce flight movements in the desired 3D trajectory.
They introduced the concept of a modular and reconfigurable framework designed for animal VR systems.
The FlyVR framework supported the configuration of multiple camera-projector systems along with accurate geometric and photometric calibration for arbitrary surfaces.
This is an advantage over previous methods as the same framework can be used with different configurations (tethered, free-flying and treadmill) for different animals.
They used open-source frameworks such as ROS and OpenSceneGraph, which are well known in the robotics and graphics community.
Additionally, they also showed a new approach where multiple flies could be tracked while the stimulus was delivered by focusing on one of the flies.

Stowers \etal\cite{stowers_virtual_2017} further extended the free moving VR systems to rodents (MouseVR) and fish (FishVR) with FreemoVR platform. 
This platform can display a wide range of stimuli, naturalistic and abstract, for experiments on multiple species in different configurations.
The FishVR system is the first underwater VR application, where visual stimuli are projected on a fish bowl from below, and infrared 3D tracking is used to for perspective correction (see figure \ref{fig:FreemovingVR}).
The study confirmed that fish responded to the virtual stimulus as if they were real. 
They avoid virtual obstacles placed in the fish tank by swimming around it as though it was present.
In addition, when a virtual conspecific (same-species fish) was introduced, they swam with them as though in the real world.
VR for freely behaving animals offers new avenues for research in the field of social and collective behavior.
The MouseVR setup is designed to allow mice to move freely on a raised circular platform (see figure \ref{fig:FreemovingVR}) where stimuli were displayed on the floor, a similar setup as used by Del Grosso \etal \cite{del_grosso_virtual_2017}.
Experiments with checkered patterns show that freely moving mice estimate height using motion parallax, a finding that was not possible to test in earlier mention treadmill based systems \cite{stowers_virtual_2017}.
Experiments with freely flying flies in FlyCave indicated that flight control of flies is fundamentally altered by tethering, even without head fixation \cite{stowers_virtual_2017}.
The authors used this study to stress the importance of developing new methods for free-moving animals. 

VR is a powerful tool for investigating mechanism of behavior.
The scope of virtual stimulation-based methods go beyond visual stimulation.
For exmaple, Sofroniew \etal \cite{sofroniew_neural_2015} studied navigation behavior in rodents using tactile stimulation \cite{sofroniew_neural_2015}, and Cushman \etal \cite{cushman_multisensory_2013} used directional speakers for acoustic stimulation.
Designing multi-sensory VR system is crucial for studying naturalistic behavior patterns, however our ability to design such methods are limited.
This paper will only focus on the limitations of VR systems with virtual stimuli.

\section{Limitations of Virtual Environments \label{Sec:Limitation}} 
In this section, we will discuss the limitations faced by researchers while designing virtual environments for studying animal behavior.
Virtual environments are designed to create believable experiences for animals by artificial stimulation of their sensory apparatus.
The main challenge is to create realistic simulations which change continuously based on the behavioral response of animals.
Currently, this is done primarily by displaying visual stimuli to the animal by using screens or projectors, and tracking their response by using a camera and treadmills.
The stimulus can also be controlled externally to introduce perturbations. 
This approach is limited to some animals because existing technology is not capable of solving tracking and simulation related challenges generally, for all species.
Most of these limitations can be attributed to the physiological properties of the animals. 
We examine the limitations of existing tracking and stimulus delivery methods and link them to the physiological properties of the animals.
Our discussions are inspired from other reviews which focus specifically on limitations of using screen/display based artificial stimuli i.e. video playback \cite{death_can_1998}, animation \cite{woo_dummies_2011} and VE \cite{chouinard-thuly_technical_2017,stowers_reverse_2014}.

\subsection{Limitations of stimulus design and delivery}
Animal vision has evolved for enhancing survival, therefore different animals have different visual properties such as color vision, the field of view, etc.
The stimulus employed must therefore be compatible with the requirements of each animal's visual system.
The stimulus design and delivery approach are also crucial for the success of behavioral experiments.
In these respects, all commercially available technology has limitations.
In the following text, we outline some important visual properties and relevant technological considerations.

\paragraph{Multispectral vision}is the ability to visualize different spectra of light.
It is also known as spectral resolution in the literature. 
Humans and some primates are trichromatic, which means that a combination of three colors (Red, Blue, and Green) is sufficient to cover the entire color spectrum seen by humans. 
In the case of animals, some are dichromatic (most mammals) or tetrachromatic (birds, reptiles), and some animals see completely different hues e.g. UV, UV with red, UV with green \cite{death_can_1998}.
Invertebrates commonly use polarized light for navigation and therefore show a preference towards it. 
Failure to reproduce such properties may affect the experiment. 

Technological Considerations: 
LCDs and projectors designed for human vision are useful for some animals with trichromatic vision and dichromatic vision.
In some cases, lighting conditions are changed \cite{stowers_reverse_2014} or color filters are used to match the colors on the display with the color perception of animals \cite{nityananda_novel_2018}(see figure \ref{fig:teaser}).
Creating realistic colors for animals with multispectral vision (e.g. UV) is difficult and should be considered when designing experiments.
Invertebrates have attractions towards some wavelengths of light, which should be considered in order to avoid unintended disruption to behavior.  

\paragraph{Flicker fusion threshold} is the threshold beyond which a flickering pattern appears continuous to the observer.
Visual information from the environment is integrated a certain time before it is experienced.
This integration time is varies in different species. 
Fast-moving animals typically have a higher flicker fusion threshold. 
A movie displayed at a frequency of 25 Hz is sufficient for humans to perceive continuous motion but for bees, the same movie would appear flickering. 
The illusion of motion may be broken by the slow or glitchy movements of the stimuli.

Technical Considerations:
Flicker fusion is important property when selecting the display and lighting for illumination of experimental arenas.
The light source may appear to flicker if animal's flicker fusion threshold is higher than operational frequency of the light source.
Potential biological affects of artificial light flicker are discussed in detail by Inger \etal \cite{inger_potential_2014}.
The same is applicable for operation frequency of the display or projector, and framerate of the rendered simulation. 
Most existing methods use displays at 120 Hz and employed GPUs for fast rendering. 
It was shown that the flicker fusion threshold of some animals can be lowered by manipulating size of stimulus, luminance of display, brightness of surrounding and region of retina involved in image formation \cite{death_can_1998}. 

\paragraph{Visual acuity} of an animal is it's ability to resolve spatial detail. 
It can also be defined as the spatial resolution of the eyes.
It is measured in degrees; some animals have very high acuity (e.g. eagles, falcons) and some have very low acuity (e.g. fish or ants) \cite{death_can_1998}.
The displayed stimuli may appear pixelated or unwanted holes can be seen in images when the acuity of animal is not appreciated.

Technical consideration: 
Visual acuity is considered when selecting the display screen or projector and the distance at which to present the stimuli. 
Animals with very high visual acuity may see pixelated images on screens made for humans. 
Similarly, distance from the screen is essential because the effect is greater at closer distances. 
If this requirement is unmet, the illusion of continuous color might be broken, which may be an important consideration for the experiment \cite{death_can_1998}. 

\paragraph{Field of view (FOV)} refers to the area/volume observed by the eyes at any given moment. It is usually measured in degrees and can vary widely between different animals.  
FOV depends on the position of the eyes and construction of the eye. 
Animals with front-facing eyes and overlapping vision (e.g. primates, cats) have considerably smaller FOV than animals with eyes on the sides of the head (e.g. birds or insects). 
FOV may change slightly for animals that can rotate their eyes in the socket. 
It is also notable that most animals do not have sharp vision in all parts of the FOV i.e. high resolution at fovea and less at the periphery. 

Technical consideration: 
FOV is considered while deciding the display area and shape of the screen for the stimulus. 
Most of the time curved or cylindrical display screens are selected for small insects and rodents \cite{thurley_virtual_2017}.
Projectors are preferred over LCD screens because curved LCDs are difficult and expensive to produce. 
A larger FOV is ctypically overed by using multiple projectors, which adds the additional complexity of synchronization and calibration of projectors.

\paragraph{Depth Perception}
Most animals have some mechanism to perceive depth in the environment. 
Different animals use different cues such as stereopsis, motion parallax or focusing, overlap, shadow, vertical distance to the horizon, retina to image size ratio, perspective and texture \cite{death_can_1998}. 
Biologically stereo vision is not always necessary for all species, many species use non-stereoscopic depth cues because they have limited overlap between field of view. 
Failure to accommodate some depth cues may reveal the 2D nature of the stimulus \cite{collett_vision:_1996,woo_dummies_2011}.

Technical consideration: 
Depth cues are considered while designing appearance of virtual stimuli and the experiment.
Until now, most artificial stimuli based methods display stimulus on flat or curved surfaces.
It is likely that animals can perceive flat or curved screen if the stimulus is rendered without correct perspective correction, which may affect their behavior.
Recent VR methods use 3D head tracking or body tracking to maintain perspective of the animal but do not offer stereoscopic depth cues. 
Researchers must be careful while designing experiments which require the animal to may be use depth cues.
Stowers \etal \cite{stowers_virtual_2017} suggest that tracking eye movements is important for introducing depth cues. 
Recently, Nityananda~\etal \cite{nityananda_insect_2016} glued color filters to study stereoscopic depth perception in insects (see figure \ref{fig:teaser}). 
They show that it is possible to use such modification when the research question is chosen appropriately. 

\subsection{Limitations of tracking}
Tracking movements and the perspective of the animal is essential for designing a VR experiment. 
Tracking freely-moving animals is challenging and most of the experiments still require tethering or other restrictions. 
Adding markers on animals may affect their natural behavior, but recent advances in computer vision have shown promising results for markerless tracking \cite{perez-escudero_idtracker:_2014}.
Existing tracking limitations often stem from physiological properties which are explained below.

\paragraph{Locomotion properties}
Animals possess diverse abilities to move in their environment, such as flying, swimming or jumping. 
Often the speed of locomotion may vary and some movements (e.g. jumping) have to be restricted in order to keep the animal in a desired space.
Movement of the animal in the arena must be measured for accurate depiction of stimuli in the virtual environment and for rendering perspective correct stimulus.
Accurate movement tracking is necessary for mapping movement decisions of the animal to the visual features rendered in the virtual world.
Mismatch between this mapping can potentially invalidate behavioral findings.

Technical Consideration:
Locomotion properties of the animal influence the selection of the tracking approach. 
Ideally, the animal should be freely moving but restriction may be necessary depending on the need of experiment (e.g. neurophysiology).
Tethering or treadmill based approaches (figure \ref{fig:FreemovingVR}) may be selected if feedback from other sensory systems can be compromised or disregarded for the purpose of the study. 
In both cases, optical tracking is used for tracking movement of the animal.
In the case of treadmills, the motion of ball is measured to infer the movement of the animal. 
The cameras selected for sampling the motion of the ball must operate at higher frame rates than the rate of rotation of the styrofoam ball.
Additionally, the rotation mechanism of the ball must sensitive towards variations in the movement of the animal.
For example, the ball must accelerate and decelerate in sync with the animal’s motion otherwise it may create an unwanted perturbation for the animal \cite{dahmen_naturalistic_2017}.
Similar considerations must be made when selecting cameras for tracking motion of the animal in tethered cases. 
For example, Stowers \etal \cite{stowers_virtual_2017} used sampling frequency of 100 Hz to compute motion of the fruit flies.   

Computer vision algorithms are used to track motion of freely moving animals. 
The performance of such algorithms is dependent on the visibility of the animals in the images. 
Fast-moving animals can appear blurry as a result of improper camera selection.
Camera properties such as e.g. frame rate, resolution, opening angle, rolling/global shutter, must be considered to capture the movements of the animal.
3D tracking requires multiple cameras which adds technical complexity regarding calibration and synchronization of cameras.
Active treadmills are designed to restrict animals to one particular spot to reduce tracking complexity. 
Often lighting configurations are selected to enable real-time tracking.
For example, IR is preferred because it is easier to add much more light to the scene without disturbing behavior of the animal \cite{stowers_reverse_2014}.

\paragraph{Appearance} of many animals differs in terms shape and structure which presents novel challenges for purely image-based tracking of animals.
Some animals do not have any conspicuous features, and some have repetitive patterns which makes different parts of the body appear confusingly similar.
Such confusions lead to fluctuations in the tracking results, and consequently in the presented stimuli.

Technical considerations: 
The appearance of the animal is an important consideration for the selection of tracking software, camera and light conditions. 
The software must process the image in real-time and detect the animal.
Paint or reflective markers can be used to add features to have seamless tracking results.
Many recent improvements in marker-less tracking methods can allow real-time tracking of seeming featureless objects.
Light conditions are often changed to increase the detection rate of markers or animals, while high-resolution cameras are useful for capturing fine details of the animals.
However, high-resolution images require a longer time for processing and storage, and therefore the selection of camera often involves a tradeoff. 

\subsection{Latency}
Latency of a closed loop system is the overall delay between movement of the animal and the change of stimulus on the display.
Multiple computational steps are involved between these two events such as image processing, data storage, graphics rendering, etc.
Each of these steps introduces a time delay in the system. 
Overall latency of the system is caused by both software and hardware components. 
The latency must be very low to allow for an interactive experiment in real-time. 
It is difficult to provide economical solutions for many hardware-related problems e.g., fast computation, higher bandwidth data transmission, and responsive displays. 
Because of these challenges, the achievable overall latency of the system must be considered while selecting the animal and the behavior to study. 

\subsection{Lack of technical expertise}
VEs for animals are designed by biologists using technology developed by engineers.
Modern VEs are designed using software and algorithmic methods from computer vision and XR communities such as tracking, graphical rendering, because their work available through open-source distribution. 
These methods are complex and certain expertise are required to tweak these methods to be able to use them with animals.
Biologists are forced to develop engineering and programming skills to develop new concepts for behavioral experiments in virtual environments.
Only few biologists have successfully bridged the gap between these fields to create customized tracking algorithms and modular software frameworks for performing experiments with animals in VE.
We consider that lack of technical expertise, from CV \& XR community, is one of the biggest limitations for the development of VR for animal behavior experiments.


\section{Discussion Of Future Directions \label{Ref:SecDicussion}}
%
The aim of this paper is to promote interdisciplinary research between engineers, computer scientists, and biologists.
In this spirit the focus of our discussion will be on the future of interactive behavior experiments.
We will report the current research trends in robotics and computer vision. Based on that we will suggest two novel applications for behavior studies, which may be realized with support from experts in the XR community.
Our intention is to provide a starting point that may encourage further discussion on this topic in the community (Ref. supplementary document for additional discussions).

Existing VEs are not yet suitable for studying all types of behaviors.
The animal's sensory feedback is generally restricted to prefer one sensory input using methods previously described such as wing clipping, body fixation.
This is not ideal as lack sensory information affects the animal's decision making process in some cases \cite{stowers_virtual_2017}.
Moreover, animals with multi-modal sensory systems combine information from multiple senses e.g. sound, vision or smell.
It is a major challenge to design multi-sensory VEs for freely moving animals.
Biologists have worked meticulously to gain knowledge about the behavior and sensory systems of some animal species.
Based on this information they have built interactive VEs for some animals e.g. insects, fish, and rodents. 
There is a strong need for the development of new solutions which will improve the sensory feedback mechanisms, add multi-sensory feedback and extend the application of VR to new species.
We argue that some of these problems can be alleviated in the future by starting new collaborations between experts in computer science and biology.

\subsection{Growing support from robotics and computer vision}
Animal behavior studies are moving towards an increasingly interdisciplinary approach.
The behaviors and mechanisms studied in the animal VR are gaining attention in the field of robotics.
Vision induced locomotion and navigation studies in small insects are appealing for designers of nature-inspired robots \cite{krause_interactive_2011,Zou_2016_robots,balasubramanian_insect-sized_2018} and self-navigating drones \cite{jafferis_untethered_2019}. 
Studies focusing on the understanding of sensory mechanisms of small animals are gaining interest in the field of smart sensor design.
In 2018, DARPA launched a robotics challenge to design small, lightweight and power-efficient micro-robots for use in disaster relief scenarios of the future.
The recent developments in free moving VR systems have opened doors for conducting new types of studies in collective behavior and social behavior, with developers of self-organizing robots already using the theories developed in collective behavior studies \cite{tero_rules_2010, werfel_designing_2014}.
The robotics community is actively involved in development of new methods for studying behavior using interactive robots \cite{klein_robots_2012,krause_interactive_2011,landgraf_robofish:_2016,webb_what_2000}.

Real-time methods for tracking of the eye positions and head orientations of animals is missing in existing VEs.
This improvement in tracking is important to extend the application of VEs to other animals.
In case of small animals the head position is inferred from the animal’s position and orientation, however, it is difficult to do the same animals with articulated bodies.
Recent publications in computer vision literature show that the community is taking interest in challenging problems involving animal tracking.
Many researchers have proposed easy to use methods for keypoint based posture computation in animals \cite{graving2019deepposekit,mathis_deeplabcut:_2018} (Fig. S2,S3 in supplementary).
Moreover, extracting 3D posture of animals from images and videos is an emerging topic in the computer vision community \cite{DeepFly3D,zuffi_lions_2018,novotny2019c3dpo,mathis_deep_2019}.
Posture based video analysis and activity recognition are currently being investigated using deep learning techniques.
High-resolution temporal information with postures may allow the study of complex behavior patterns e.g. courtship display or aggression display.

\subsection{Introducing new concepts of XR \label{subsec:XR}}
\paragraph{Spatial Augmented Reality (SAR)} applications are not fully explored in animal behavior experiments.
Projectors are readily used in behavioral experiments but often their use is limited to displaying stimuli in open-loop e.g. predator-pray interaction \cite{Ioannou1212}.
Ioannou ~\etal \cite{Ioannou1212} used open-loop approach due to the lack of methods to track fish in real-time.
Now, it is possible to perform similar experiments in closed loop with the help of real-time tracking methods. 
One possible application is the use of dynamic projection mapping with robotic animals. 
It was shown that social behavior can be studied using robots instead of real animals, and animals can and do interact with robots \cite{landgraf_robofish:_2016, krause_interactive_2011}.  
Ladgraf~\etal \cite{landgraf_robofish:_2016} tested robotic models with different appearances and claimed that appearance was crucial for the acceptance of a robotic agent by real fish to study the social behavior of fish.
We argue that dynamic projection mapping techniques \cite{Narita_SAR_2017,Miyashita_SAR_2018} can be deployed to alter features of robotic stimuli.
The projector can project different patterns on a robotic agent while maintaining the perspective of the real animal using real-time 3D tracking.
Experiments with SAR could open new possibilities such as training animals for navigation or memory experiments using virtual agents projected on a wall or a robot. 
Researchers at CASCB \footnote{Centre for the Advanced Study of Collective Behavior, Konstanz} are currently building a large scale arena for conducting interactive experiments with one or more animals using real-time 3D tracking and projection technology.

\paragraph{Collective behavior studies} related to the decision making of a group and the effect of the individual decision on the group may be studied using VR.
It is shown that real animals do interact with virtual conspecifics in the VR e.g. fish\cite{stowers_virtual_2017}.
Multiple VR systems can be plugged together to create conditions for social behavior in a virtual manner i.e. collaborative VR space for animals \cite{larsch_biological_2018}. 
In such a virtual social scenario, each animal may interact with a group of virtual conspecifics which are projections of real animals from other VR systems.
The visual information available to each individual can be controlled in such environments and manipulated based on the needs of the experiment. 
If the animals start to swarm in virtual environments the principal’s governing their decisions can be studied in much detail. 

\section{Conclusion}
Applications of virtual environments have previously been discussed from the perspective of a human user. 
However, there is substantial work showing that some animals can, and do, respond to virtual stimuli.
In this review, we discussed the use of the virtual environments for studying animal behavior.
We show that investing in the development of such concepts is beneficial for research in various disciplines throughout biology and engineering.
A lot of progress has been made in the past two decades, but support in terms of technology development is required to extend the use of this technology in biology.
We hope that this review sparks interest in animal oriented applications of VE among the technology developers in the XR community.
Animal behavior experiments involving XR systems and robotics have the potential to become an independent field of interdisciplinary research.
\acknowledgments{
The authors wish to thank John Stowers, Yuji Oyamada, and Bianca Schell for reviewing the manuscript. Authors who provided their images. This work is support by funding from the DFG Centre of Excellence 2117 “Centre for the Advanced Study of Collective Behaviour" (ID: 422037984).}

\bibliographystyle{abbrv-doi}

\bibliography{Zotero,template}

\end{document}